\documentclass[conference,pageno,bookmarks=false,breaklinks=true,colorlinks,linkcolor=black,citecolor=blue,urlcolor=black]{IEEETran}

\usepackage[normalem]{ulem}
\usepackage{makecell}
\usepackage{cite}
\usepackage{amsmath,amssymb,amsfonts}
\usepackage{mathptmx} 
\usepackage{comment}
\usepackage{bm}
\usepackage{algpseudocode}
\usepackage{algorithm, algorithmicx}%
\usepackage{graphicx}
\usepackage{textcomp}
\usepackage{multirow}
\usepackage{fancyhdr}

\title{SafeBet: Secure, Simple, and Fast Speculative Execution}
\author{
Conor Green*, Cole Nelson*, Mithuna Thottethodi, and T. N. Vijaykumar\\
Elmore Family School of Electrical and Computer Engineering, Purdue University\\
\{green456,nelso345,mithuna,vijay\}@purdue.edu\\
*Equal contributors
}

\date{}
\usepackage{ifthen}
\usepackage{calc}
\usepackage{pifont}
\usepackage{color}
\usepackage{fancyhdr}
\usepackage{xspace}
\usepackage{url}
\usepackage{xkeyval}
\usepackage{multirow}
\usepackage[table]{xcolor} 
\definecolor{lightgray}{gray}{0.9}
\newenvironment{stripetabular}{\rowcolors{2}{white}{lightgray}\tabular}{\endtabular}

\newenvironment{stripetabular2}{\rowcolors{3}{white}{lightgray}\tabular}{\endtabular}

\newcounter{hours}
\newcounter{minutes}

\newcommand{\JS}{{JavaScript}\xspace}
\newcommand{\name}{{SafeBet}\xspace}

\newboolean{timeofmake}
\setboolean{timeofmake}{false}

\newcommand{\dontinclude}[1]{ }

\newcommand{\putsec}[2]{\vspace{-0.1in}\section{#2}\label{sec:#1}\vspace{0.0in}}
\newcommand{\putsubsec}[2]{\vspace{-0.05in}\subsection{#2}\label{sec:#1}\vspace{0.0in}}

\newcommand{\tabput}[3]{
\begin{table}[t]
\caption{#3 \label{tab:#1}}
\vspace{-0.1in}
\begin{center}
{
#2
}
\end{center}
\vspace{-0.15in}
\end{table}
}

\newcommand{\tabputW}[3]{
\begin{table*}
\caption{#3 \label{tab:#1}}

\begin{center}
{
#2
}
\end{center}

\end{table*}
}

\newcommand{\figput}[4][1.0\linewidth]{
\begin{figure}[t]
\begin{minipage}{\linewidth}
\footnotesize 
\begin{center}
\includegraphics[width=#1]{figures/#2}
\end{center}
\vspace{-0.2in}

\caption{#4 \label{fig:#2}}

\vspace{-0.2in}

\end{minipage}
\end{figure}
}

\newcommand{\figputW}[4][\linewidth]{
\begin{figure*}
\begin{minipage}{\linewidth}
\footnotesize 
\begin{center}
\includegraphics[width=#1]{figures/#2}
\end{center}
\vspace{-0.2in}
\caption{#4 \label{fig:#2}}
\end{minipage}

\end{figure*}
}

\newcommand{\figref}[1]{Figure~\ref{fig:#1}}
\newcommand{\tabref}[1]{Table~\ref{tab:#1}}
\newcommand{\secref}[1]{Section~\ref{sec:#1}}

\newcommand*\rot{\rotatebox{90}}

\begin{document}

\maketitle
\thispagestyle{empty}

\pagestyle{plain}

\begin{abstract} 

Spectre attacks exploit microprocessor speculative execution to read and transmit forbidden data outside the attacker's trust domain and sandbox.  
Recent hardware schemes allow potentially-unsafe speculative accesses but prevent the secret's transmission by delaying all or many of the access-dependent instructions {\em even in the predominantly-common, no-attack case}, which incurs performance loss and hardware complexity. 
Instead, we propose \name which allows {\em only}, and  {\em in the common case} does not delay {\em most}, safe accesses. We first consider the simple case
without (i) {\em code} control flow transfer across trust domains  or (ii) {\em data memory} sandbox boundary changes (i.e., some locations dynamically move across sandboxes);
and then extend to the complex cases with these behaviors. 
We make the key observation that in the simple case 
speculatively accessing a {\em destination} location  is safe if the location's access by the same static trust domain has been committed previously (i.e., the domain's instructions are permitted to access the location); and potentially unsafe, otherwise. The domain's code memory region is called the {\em source}.
\name employs the Speculative Memory Access Control Table (SMACT)  to track non-speculative source region-destination address pairs.
Disallowed accesses wait until reaching commit to trigger well-known replay without any intrusive hardware changes.
Unlike the simple case, however,
inter-trust domain control flow allows the permissions obtained legitimately by one 
dynamic execution instance to be exploited by another instance to access forbidden data. 
To prevent this loophole,  \name replaces the static source in each permission with a unique instance identifier.  Further, the permissions are revoked when a trust domain's sandbox boundary changes dynamically to prevent the use of stale permissions. 
\name exploits the redundancy in the destination address bits to shrink the SMACT using bit masks. Finally, to enlarge the effective SMACT capacity, \name safely coarsens the destination granularity.
\name prevents  all variants of Spectre and Meltdown except Lazy-FP-restore, 
based on any current or future side channel while using only simple table-based access control and cache miss replay with virtually no change to the pipeline.
Software simulations show that \name uses an 8.3-KB SMACT per core to perform within 6\% on average (63\% at worst) of the unsafe baseline behind which {\em NDA-restrictive}, a previous scheme of security and hardware complexity comparable to \name's, lags by 83\% on average.

\end{abstract}

\putsec{intro}{Introduction}
Speculative execution-based security attacks, called Meltdown ~\cite{meltdown,markhill-slides} and Spectre~\cite{spectre,markhill-slides}, affect most  high-performance microprocessor-based computer systems.
These serious, hardware-based attacks breach {\em trust domain} and {\em sandbox} boundaries to read and transmit the kernel or browser memory at viable rates (e.g., 10s-100s KB/s). While many software-based attacks exist (e.g., buffer overflow), these hardware-based attacks have significantly enlarged the attack surface.

Fundamentally, these attacks exploit speculative execution, a key performance feature of modern microprocessors. These attacks exploit the facts that   (1) incorrect execution before
misspeculation detection can be leveraged to {\em
access transiently and illegally} secrets outside the attacker's trust domain and sandbox, 
and (2) upon detecting misspeculation,
modern architectures clean up the architectural state  (e.g., register and memory) but not the micro-architectural state (e.g., branch predictors and caches). The {\em surviving} micro-architectural state can act as a side channel for transmission. In contrast, legally holding secrets and being tricked into leaking them is a privacy problem, not our focus (\secref{related}). 

There are three main Spectre variants which include many sub-variants~\cite{canella} (See~\tabref{threat}). The first variant (e.g., CVE-2017-5753) circumvents software bounds-checking  by exploiting branch prediction to  transiently load forbidden data 
(e.g., JavaScript accessing  the Web browser's data). 
The second variant  (e.g., CVE-2017-5715)  injects an indirect branch target (or return address~\cite{spectre-RSB}) from the attack process 
to exploit  a \textit{gadget} (i.e., an attacker-selected  code snippet) in a victim process to transiently load  forbidden data (e.g., a user process fooling the kernel). The final variant (e.g., CVE-2018-3639), known as Spectre-v4 (Meltdown is Spectre-v3),  exploits speculative store bypass.
We target all variants and sub-variants of Spectre and Meltdown; we do not target classic, non-speculation-based microarchitectural side channel attacks~\cite{brpred-leak1,brpred-leak2,flush+reload,prime+abort,sandbox-spy,LLC-practical,TLB-leak,TLB-leak2,FPU,membus}.

Completely disabling speculative execution, which avoids long
latencies via highly-accurate prediction,
would result in unacceptable performance loss. 
The possibility of the attacks does not remove the need for speculation. Our guiding principle is that 
\textit{while a forbidden access is known to occur in only one way (speculatively), 
the secret can be transmitted through numerous current and future micro-architectural side channels}~\cite{flush-reload1,flush-reload2,prime-probe,evict-reload,flush-flush,evict-time,dram-channel,tlb-channel,pagecache-channel,spectre-prime,netspectre,smotherspectre,uopcache}. Approaches to plug the transmissions 
must cover \textit{all} channels  which is harder than preventing the forbidden  access. In fact,  rolling-back micro-architectural state to plug the transmission~\cite{cleanup} may be susceptible to timing channels~\cite{spectre}.   Some proposals ~\cite{invisispec,cond-spec,cfence,index-encrypt,delayonmiss}
plug specific side channels but make invasive hardware changes (e.g., changes to cache coherence), incur performance loss,  or remain susceptible to other side channels discovered later. 
Other proposals allow potentially-unsafe accesses  but  prevent the secret's  transmission in order to  block all side channels  by  delaying  the access-dependent instructions until the access is  no longer speculative  ~\cite{NDA,spec-shield,spectaint,data-oblivious}.
Unfortunately, these schemes require complex hardware with considerable feasibility challenges or incur significant performance loss (\secref{related}).

Rather than
allowing a potentially-unsafe speculative access and then blocking its transmission to  all~\cite{NDA} or many~\cite{spectaint, data-oblivious} dependent instructions  {\em even in the common, no-attack case}, we propose \textit{\name} which allows {\em only}, and {\em in the common case} does not stall {\em most}, safe accesses, foreclosing any possibility of transmitting the secret. \name protects  both (a)  the simple  case where 
{\em code} control flow transfer across trust domains (e.g., due to shared library calls or attacker-selected victim code snippets
called gadgets)
and {\em data memory} sandbox boundary changes (e.g., via  memory de-allocation, software access-control modification, or recompilation) are absent;
and  (b) the complex cases which include these behaviors. 
\name is  based on  our key observation that in the simple case,  \textit{speculatively
accessing a location is safe if the location's access by the same static trust domain has been committed previously (i.e., the domain's instructions  are permitted to access the location)}.
The domain's code {\em region} (e.g., of GB granularity where the 
domain's code is placed aligned to the granularity)
and the location are  called the \textit{source} and \textit{destination}, respectively. 
The permission is per source region 
because some destinations are  forbidden for some source regions within the same process  in some attack scenarios (e.g., the browser data is forbidden for the tabs within the browser process). 
A data  access deemed potentially unsafe can proceed only upon reaching commit.
Thus, while no unsafe access  (false positive) is allowed, a few  safe accesses may  be delayed (false negative), resulting in some modest performance loss.

We make the following contributions. First, to capture the simple case, we propose to track \textit{non-speculative data access  source-destination pairs} in the \textit{Speculative Memory Access Control Table (SMACT)}~\cite{smact-patent}. 
While the permissions  are created in the SMACT upon the accessing instructions reaching commit, 
speculative instructions look up the SMACT in parallel with the data cache. 
Disallowed accesses wait  until the instruction reaches commit. Instead, waiting only until the instruction becomes non-speculative would be faster which is important when all or many accesses are delayed, but requires complex hardware changes
~\cite{NDA,spectaint}. Because  \name's false negative rates,  due to misses in the SMACT, are low,
\name employs the simpler choice by leveraging the well-known \textit{cache miss replay}. 

Second, unlike the simple case, attacks can abuse inter-trust domain function calls to shared functions to  transiently access forbidden data  that the function had previously accessed non-speculatively in another call (including invoking the victim and forcing a speculative access followed by a leak, as in Spectre-v2). The shared function is a ``Confused Deputy'' in this attack~\cite{Confused-Deputy}.  For example, a browser-tab shared  utility function  called from the browser may acquire permissions to the browser data which a malicious tab could access later via the same function's permissions. 
To prevent this loop hole, \name replaces the static source in  the permission with a unique \textit{instance identifier} so that permissions for the same static source region in one instance cannot be used in other instances {\em with or without} context switches~\cite{instance-patent}.

Third, a source-destination pair's permission must be revoked whenever the sandbox boundary changes -- the other dynamic behavior; otherwise, stale permissions could allow illegal accesses to locations that are no longer
within the sandbox. 
Because hardware invalidations for such revocations in a multicore may incur cache coherence-like complexity, we employ software revocations, similar to TLB shootdowns. 
To amortize the software overhead imposed by frequent freeing (e.g., once per 22K-70K instructions in some SPEC benchmarks), we {\em lazily} batch  several frees based on the key observation that any frequent freeing in applications is typically  of small chunks of memory (e.g., 64-128 B). Consequently, the delayed reclamations due to our batching does not significantly impact memory footprint or performance. While batching TLB shootdowns raises correctness issues~\cite{LATR,TLBshootdown}, batching frees affects only performance  but not correctness.

Fourth, using fewer destination address bits than needed to shrink the SMACT would induce potentially unsafe aliases where the permissions of a  source-destination pair would include  the other aliases. We safely reduce these overheads by exploiting the redundancy in the upper-order address bits via bit masks.

Finally, to increase the effective capacity and decrease the false-negative rates of the SMACT, we coarsen the granularity of the destination.  However,  avoiding aliases requires that the minimum size for dynamic memory allocation match the coarse granularity (e.g.,  no larger than 32-64 B to avoid  fragmentation).

\name prevents all variants of Spectre and Meltdown except Lazy-FP-restore, using any current or future side channel. 

Software simulations show that compared to the unsafe baseline,
\name uses 8.3 KB per core for the SMACT to perform within 6\% on average (and 63\% at worst), whereas {\em NDA-restrictive}, a prior scheme  of security and hardware complexity comparable
to \name's, is slower than the baseline by 83\% on average. Yet, \name leverages simple 
table-based access control and replay with virtually no pipeline change.  

\putsec{background}{Threat Model}
As explained in~\secref{intro}, we focus on speculation-based Spectre variants~\cite{spectre,netspectre,sgx-spectre,ssb,spectre-RSB,ret2spec}. \name also covers Meltdown variants~\cite{foreshadow,meltdown,ridl,fallout,foreshadow-ng}.
\tabref{threat} summarizes \name's threat model and coverage.

\tabput{threat}{
\begin{stripetabular2}{p{0.8in}cc}
\bf Access &  \bf Attack & \bf \name\\
\bf Mechanism &  \bf example & \bf coverage\\
\hline
Unsafe Load & \cite{spectre,spectrev1.1} & SMACT  \\
Bounds Store & \cite{spectrev1.1}  & SMACT  \\
Store Bypass & \cite{canella,ssb} cross trust domains~\cite{msr-ssb-trust} & Instance \\
Unsafe Return & \cite{spectre-RSB,ret2spec} &  Instance \\
Unsafe Branch &  \cite{spectre,smotherspectre,sgx-spectre,sgxPectre,spectrev1.1} & Instance \\
Homonym forw. & Meltdown-P~\cite{ridl,fallout,foreshadow,foreshadow-ng} & SMACT \\
Meltdown & BR\cite{canella}, GP\cite{rsre}, RW\cite{spectrev1.1}, US\cite{meltdown}  & SMACT \\
Protection keys & Meltdown-PK~\cite{canella} & Revocation \\
FP registers & Lazy-FP-Restore~\cite{LazyFP} & No\\
Non-transient & \cite{NDA,uopcache} & No\\
\hline
\end{stripetabular2}

\vspace{-0.2in}
}
{Threat model and \name's coverage}

\putsubsec{spectre}{Transient-access attacks}

In general, the Spectre attacks access secret data under mispredicted wrong-path instructions -- typically due to control-flow misspeculation -- and transmit the secrets via microarchitectural side channels. In contrast, the Meltdown variants exploit speculative execution past exceptions/faults~\cite{canella}. 

For example, Spectre-v1  bypasses  bounds checks in contexts where some external code is run \textit{within} a ``host'' process  (e.g., downloaded \JS  within the browser or packet filter bytecode within the kernel~\cite{bpf}). These contexts choose within-process sandboxing instead of virtual memory isolation of separate processes for performance. Further, language features (e.g., no pointer arithmetic in \JS~\cite{ecma}) ensure that the bounds checks are sufficient for safety. 

In Spectre-v1, the attack mistrains the branch predictor to predict a bounds-checking branch to take the within-bounds path (Alg.~\ref{alg:spectreattack}). Then the attack provides an  invalid input  
causing a misprediction, so that a bounds-check-protected load  makes an   out-of-bounds, {\em transient} access of a secret. A subsequent, dependent access leaves a trace of the secret in the previously-primed data cache. After the misspeculation is handled, the attack probes the cache to observe the trace and learn the secret. 
The hallmark of Spectre-v1 is the transient access of the secret which the attacker {\em cannot} access  non-speculatively. 

\begin{algorithm}
\caption{\label{alg:spectreattack}Bypassing Bounds Check (Spectre-v1)}
\begin{algorithmic}[1]
\State indexArray = [0,1,2,3,4,{\textcolor{red}{1048557}}] \Comment Mistrain + \textcolor{red}{attack}
\For{$i \in indexArray$}
    \State \textcolor{blue}{\bf t1} $\gets$ X[i] \Comment Load array element (and \textcolor{red}{secret})
    \State t2 $\gets$ L[c*\textcolor{blue}{\bf t1}] \Comment Transmit secret via data-dependent load
\EndFor
\end{algorithmic}
\end{algorithm}

In Spectre-v2~\cite{spectre} and Spectre-RSB~\cite{spectre-RSB} cross-process attacks, the attacker process primes the branch target buffer (BTB) and return address stack (RAS) to point to a gadget. The victim process is fooled into branching to the gadget which {\em transiently} accesses secrets. SgxPectre~\cite{sgx-spectre}  breaches SGX enclaves running in the same address space as the victim process via  Spectre-v2. 
Other mitigations for Spectre-v2 prevent such cross-process and cross-domain (e.g., user-kernel) attacks by flushing the BTB and RAS upon context switches~\cite{intel-spectre-whitepaper},  partitioning the resources among SMT contexts, or using process identifiers (e.g., PIDs). 
 
Spectre-v4,  which has not been shown to be practical, leverages loads that speculatively bypass stores and read stale values (e.g., overwritten secrets). However, in addition to the speculation, such loads must also speculatively cross a trust-boundary~\cite{msr-ssb-trust} and speculatively load a non-speculatively inaccessible address (otherwise, the attacker could directly read the secret).
\name prevents unsafe speculative accesses, thereby blocking all Spectre-v1, -v2, and -v4 variants independent of the side channel for transmitting secrets. 

In general, Meltdown variants access prohibited memory locations (and some registers) using transient instructions while faults/exceptions are handled lazily. These variants include user processes accessing kernel memory~\cite{meltdown},  SGX enclave data~\cite{foreshadow}, special registers~\cite{rsre}, and FP registers~\cite{LazyFP}. 
In all these cases except Lazy-FP-restore~\cite{LazyFP} (explained in~\secref{other}), \name blocks the unsafe accesses.

Some Meltdown variants~\cite{foreshadow-ng,fallout,ridl} are Intel-specific, arising from speculative accesses under exceptions, faults or TLB misses.  Such aggressive speculation can forward the data of homonyms,
which may be chosen inside the attacker's address space~\cite{fallout} or be created maliciously by hostile guest VMs~\cite{foreshadow-ng}. 
Some attacks forward data based on just partial address matches ~\cite{fallout}. 
The root of the problem is Intel's lazy checking of the TLB or not checking at all upon TLB misses while speculatively supplying the data across protection domains even for benign accesses (i.e., accesses to legitimate locations in one's own domain).
Non-Intel architectures perform eager checking and do not speculate past TLB misses.
With slight modifications to the default design, \name can cover these cases (\secref{homonym}). By delaying dependent wake-up of all accesses, NDA and Speculative Taint Tracking~\cite{spectaint} also prevent these homonym-based attacks.

Invisispec's~\cite{invisispec} {\em Futuristic} model considers transient instructions due to load-store dependencies, memory consistency violations, and exceptions.  
By checking {\em all} transient accesses, 
\name covers this model.

\putsubsec{exclusions}{Exclusions}

\name  excludes threat models in which the secret is {\em non-transiently} present in a general-purpose register (e.g., NDA's {\em Strict} mode).
NDA acknowledges that no such attack is known to exist; 
but requires that {\em all} instructions wait till commit to wake up dependent instructions (i.e., no out-of-order issue, speculation {\em or}  pipelining of the processor backend for {\em most} instructions). 

Another work~\cite{uopcache} demonstrates the $\mu$op cache to be a powerful side channel.
The paper claims that one of its transient attacks on the pipeline frontend cannot be prevented by any of the previous Spectre mitigations which are based on the backend, as is \name. However, the attack first non-transiently accesses  the secret which is accessed transiently again later; the later, transient access is dependent implicitly  on and impossible without the first, non-speculative access. As such, this  non-speculation-based attack 
is not targeted by previous Spectre mitigations or \name.

\putsec{spec-sandbox}{SafeBet}

Recall from~\secref{intro} that instead of potentially unsafely accessing the secret and then preventing its transmission to all or most dependent instructions even in the common, no-attack case,
\name allows only and in the common case does not stall most safe accesses, preempting any transmission of the secret.
We start with the simple case where {\em code} control flow transfer across
trust domains (e.g., due to shared library calls or gadgets) or dynamically changing {\em data memory} sandbox boundaries (e.g., due to memory de-allocation, software access-control modification, or re-compilation)  are absent. In this case,
\name permits speculative accesses to a location only if the static trust domain, whose code memory region is called the {\em source},
has non-speculatively accessed the {\em destination} location in the past. To handle the complex cases with the inter-trust domain control flow and sandbox boundary changes, we later add dynamic instance annotation of the permissions and permission revocation, respectively. 

\putsubsec{spec-load}{Speculative data accesses in the simple case}

At a high level, data accesses in the simple case create source-destination permission pairs in the Speculative Memory Access Control Table (SMACT) when the accessing instruction reaches commit.
Later, dynamic, speculative source-destination pairs use the permissions until eviction from the SMACT or revocation due to a sandbox boundary change.
We first describe a virtual address-based default design which we modify later to cover the Intel-specific homonym-based attacks.

Speculative data accesses involve an access-control check in the SMACT and delayed execution at commit, if deemed potentially unsafe. For the first part, given the instruction's source region and destination address pair (both virtual addresses), the SMACT is looked up in parallel with the TLB and the data cache.
If the SMACT permits the access  (the common case), which then  proceeds as usual (our first contribution). Otherwise,  the SMACT triggers a replay similar to that for a cache miss. 
The replay occurs when the instruction reaches commit (i.e., the reorder buffer head) at which time the SMACT is updated with the source-destination pair to create a permission, possibly replacing an existing entry. A disallowed instruction may be squashed due to misspeculation, before any replay.

Unlike loads, stores are performed in the cache only upon reaching commit. Therefore, stores can bypass the SMACT even though stores may speculatively prefetch the cache block (including coherence permissions).
Note that speculative store-based attacks (e.g.,~\cite{DOLMA, spectre-prime}) use the store to leak and not to access; \name prevents the access making the leak irrelevant. Alternately, Spectre-v1.1~\cite{spectrev1.1} exploits speculative store-load bypass, instead of branch prediction, to circumvent the sandbox bounds check and then speculatively loads the secret which is prevented by \name.  
In fact, even SMACT-missing loads  can
issue  cache/TLB miss fills, and update the cache/TLB replacement metadata.  Such a  fill uses the potential secret's address which may be known to the attacker anyway and is not a secret; only the secret's value is a secret (this fill is for the data access and not the second, secret-dependent, potential leak). 
As such, the fill  places any secret value only in the cache and not the pipeline. 
The fill overlaps the miss, but not the miss-dependent instructions, with the delay until replay. 
However, store-value bypass for matching loads via the load-store queue brings the load value into the pipeline. Therefore, an SMACT-missing load cannot return such a value and must wait for replay.  

Instead of waiting until commit, the replay could be triggered sooner  when the instruction becomes non-speculative. However, detecting  this condition requires complex hardware to ensure  that all previous speculations in program order have been resolved (e.g.,~\cite{NDA,spectaint,delayonmiss}),  as explained in~\secref{related}. Irrespective of the instruction waiting until commit or becoming 
non-speculative, the dependent instructions cannot issue until replay. Fortunately, \name  incurs this overhead only infrequently (i.e., upon SMACT misses). Consequently, \name can afford the simpler option of the instruction waiting  until commit. 

Instead of the source being at instruction granularity, it is more meaningful for security that the source be coarsened to a memory  region granularity (e.g., 1 GB) if the code and static data from different trust domains within a  process are placed at different regions at region-aligned boundaries (e.g., user code, libraries either together or in subsets, browser, and kernel).
Any secret accessible non-speculatively by {\em any} part of a trust domain's code is accessible non-speculatively by {\em all} of the domain's code (i.e., speculative access permissions for one instruction can be used by other instructions within one trust domain)  
and needs protection only from other domains. 
Thus, our placement requirement ensures safety for  source coarsening. Incidentally, such coarsening (1) drastically reduces {\em locality splintering} of a destination accessed by different instructions within the same region tracked in different
SMACT entries each of which incurs separate SMACT misses,
and (2) improves effective SMACT capacity by combining multiple instructions into one coarsened SMACT entry.  
 
\figput{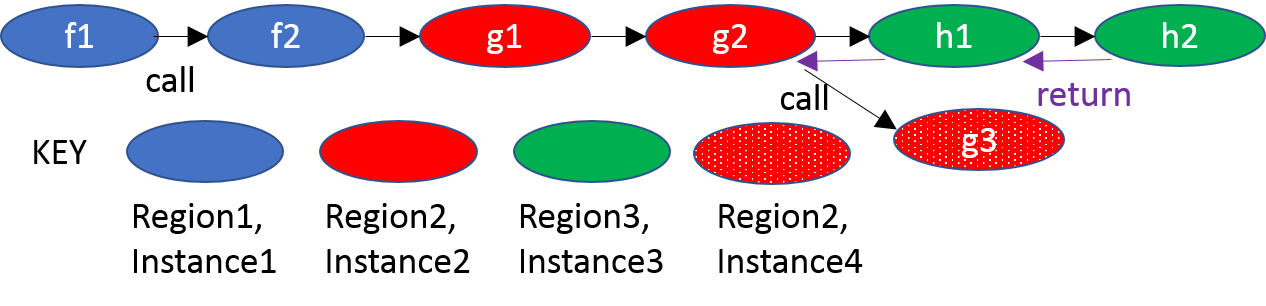}{}{Regions, calls, and instances}

\putsubsec{context}{Modifying permissions with dynamic instances}

In the absence of inter-region control flow as in the simple case in~\secref{spec-load}, 
no reuse of permissions is possible across regions whose static addresses are different. Otherwise, static addresses are insufficient. Consider a browser providing an encryption utility using a secret key where the utility is in the browser's region and is used by the tabs, each of which is in a different region.
Creating a copy of the utility for each  tab avoids sharing the utility across the tabs, but  not browser-tab sharing which, unfortunately, cannot be avoided or disallowed. 
When called legitimately from a tab (or the browser),  the utility would obtain permissions to access the key. Another, malicious tab can call the utility 
a second time  to make forbidden accesses to the key using the  permissions.  The utility is a ``Confused Deputy'' in this attack~\cite{Confused-Deputy}.

One idea is to use 
the  calling contexts to distinguish among 
calls by different regions. However, the two calling contexts in the above example are identical.
As such, the only safe option may be to create a unique instance upon every region {\em transition}, irrespective of the calling  context, to prevent permission reuse across different regions', or even the same region's, dynamic instances (e.g., different calls to a shared library). 
A key point is that any attacker calling a shared library in a victim {\em must} cross
{\em at least} the attacker's own region, creating a new instance which cannot reuse the old permissions ({\em irrespective of whether the shared library is part of the victim's region or a region of its own}).

A region $R$'s {\em instance} $I$ starts at the first call to a function $F$  in $R$ from a different region (e.g., \textit{Region2's g1} in~\figref{instance}). The duration of $I$ includes  all the calls to and returns from  functions within $R$ until $F$ returns, {\em without any}  intervening calls to and returns from functions in other regions, all of which are excluded from $I$. In~\figref{instance}, \textit{Instance2} includes \textit{g1} and {\em g2}, but not \textit{Region3}'s \textit{h1 and h2} which are in \textit{Instance3}. Because \textit{Region2}'s \textit{g3} is called after returning from \textit{Region3}, {\em g3} is in \textit{Instance4}
instead of \textit{Instance2}.
Each instance receives a unique {\em instance identifier (instID)}, (our second contribution). In our above example, a tab cannot read the browser's data (e.g., the key) or the other tabs' data (protected by the SMACT). 
Each call from a tab  to the utility is a separate instance that can read the key but the caller tab cannot read the key or the other tabs' data (protected by the instIDs). An alternative to using the InstIDs is to invalidate the entire SMACT upon an inter-region control flow. However, that option would eliminate two  key optimizations described next. 

The unique instIDs  ensure security but may degrade performance by disallowing reuse of permissions across instances. That is, every instance must obtain its own permissions, splintering
locality (\secref{spec-load}). 
Of course, the instance's later accesses can proceed speculatively, as before. 
To avoid unnecessary splintering in two common cases, we distinguish between the {\em owner} region  
and other {\em visitor} regions (e.g., the browser  or the OS is the owner, and the tabs or packet filters in Berkeley EBPF~\cite{bpf} are  visitors). 
In the case of multiple owners (e.g., browser, guest OS,
hypervisor, and host OS),  there is a hierarchical and dynamic aspect to owner-visitor relationship (e.g.,
the browser is the owner for the tabs but a visitor for
the guest OS). Fortunately, this aspect does not matter for our purposes as we see below.
The owner may provide utilities that a visitor can call (like our encryption utility example above).
We note that in {\em any} sandboxing (underlying not only \name but also NDA, STT, and others), the owner is trusted by the visitors and can access any location of any visitor.
In  the case of one owner, we assume that the loader sets the owner register.
Multiple owners can be supported by a few registers or via software convention of loading the owners at a specific address range  so hardware can identify (e.g, using upper-order address bits).  

\figput{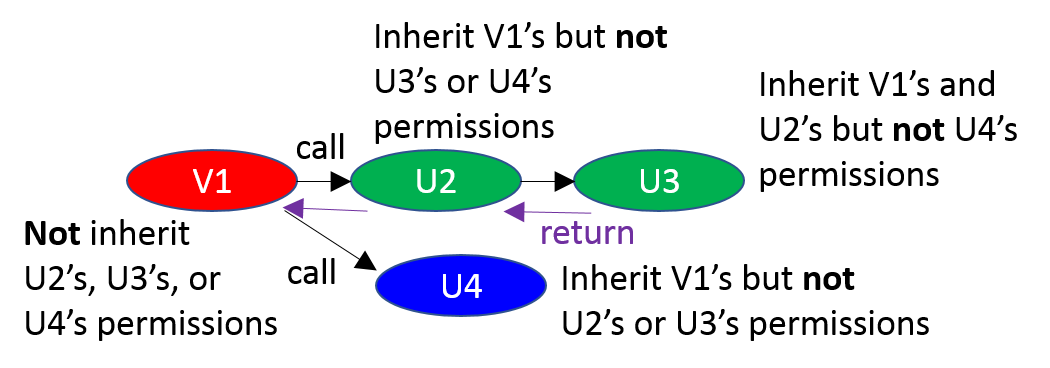}{}{Permissions inheritance}

In our first optimization, a callee instance not being able to reuse the caller's permissions causes considerable locality splintering. 
To address this issue, the trusted owner-provided utility callee instance can {\em inherit} and {\em reuse} the permissions of its visitor-caller instance, which is a common inter-region transition (e.g., U2 inherits V1's permissions in~\figref{inheritance}).
This inheritance allows the owner-callee to reuse the visitor-caller's permissions, {\em without modifying}  the permissions or {\em adding} any new permissions. 
However, this inheritance is {\em strictly asymmetric}: 
the visitor-caller cannot reuse the permissions of the utility-callee upon return; nor can multiple utility-callee instances reuse each other's permissions. In~\figref{inheritance}, V1 and U4, respectively, cannot inherit the permissions of U2, U3, or U4, and U2 or U3. 
Visitor-to-owner and owner-to-owner inheritance are allowed whereas (a) visitor-to-visitor  and (b) owner-to-visitor cases are disallowed (\tabref{call-return}).
In owner-to-owner calls, the 
caller acts as a visitor similar to valid  inheritance (e.g., browser making a syscall to the OS); the reverse act is not practical (e.g., the OS does not call a process; and
OS signal or browser call-back delivery, respectively, is {\em not} a OS-to-process or browser-to-tab call). Thus, multiple owners do not matter for inheritance purposes.
In case (a), while it is not meaningful for visitors to provide utilities for each other,  EBPF allows one filter to call another via {\em filter chaining}. In this case, one visitor should not inherit from another. Case (b), which is the reverse act, is not practical. Allowing inheritance {\em only} if the callee is the owner correctly excludes cases (a) and (b).

Second, a return to the  caller's instance is another common inter-region transition. Fortunately,  the instance is not new and 
cannot reuse the permissions of its returned callee, its own caller, or any of its own previous instances.
Therefore, upon a return, the caller instance could retain its instID for later instructions.
However, if the above reverse act were to occur (i.e.,  
the owner calls a visitor), a malicious visitor may  poison the BTB and register values (not protected by sandboxing) to reuse the owner's permissions via a gadget upon return to the owner; whereas a visitor calling  the trusted  owner  does not have this problem (\tabref{call-return}). Consequently, retaining the instID upon returning {\em only from} the owner correctly excludes the unwanted cases which start a new instance and discontinue inheritance by purging the stack (\tabref{call-return}). 

\tabput{call-return}{
\begin{stripetabular}{p{0.9in}cc}
\bf Call/return &  \bf New instance? & \bf  Inherit?\\
\hline
Call to owner & Yes & Yes \\
All other calls & Yes & No \\
Return from owner & No (retain) & No \\
All other returns & Yes & No (purge stack)\\
\hline
\end{stripetabular}
\vspace{-0.2in}
}
{Inter-region call/return}

\putsubsec{region-stack}{Implementing instances}

\name employs, in hardware, the {\em instance counter} to generate unique instIDs upon region transitions and the {\em instance stack}  to track inter-region calls and returns.
Additionally, \name maintains (1) a redundant counter, called the
{\em shadow counter}, as a safety measure for returns, as explained below; and (2) an {\em owner region register} to prevent any inheritance whenever the owner is the caller. 

The instance stack is similar to the call stack with a key difference: only the calls or returns that cross regions {\em with or without} context switches (i.e., the PC and target are in different
regions)  push to or pop from the stack.
Calls or returns within a region do not modify the stack.  A region-crossing call increments the instance counter to generate
a new instID,  pushes the target's region and the new instID on the stack, and also increments the shadow counter.  An inter-region return pops the stack.
To ensure unique instIDs, the return decrements the shadow counter but  {\em increments} the instance counter. 
However, any call or return, including an inter-region one, is typically fetched under speculation which may later lead to a squash. 
To avoid repairing the instance stack upon squashes, we push to  or pop from the stack only when the call or return
reaches commit.  The instance increment occurs speculatively at decode, decoupled from the stack operation. In the time window between the fetch and commit of the call or return, any access from the new instID would not match the instID on the top of the stack (TOS)  and 
waits until commit. This waiting occurs only at inter-region transitions and not within regions. (Speculatively updating the stack and upon a squash repairing the stack via checkpointing, similar to rename tables, is an option.)

Because the instID uniquely identifies the current instance of the source region (i.e., the instID subsumes the source region), 
the SMACT now {\em replaces} the source with the instID. An SMACT lookup matches the destination tags  of the entry and access, and the instIDs of the entry  and {\em either} the access {\em or} one level below the TOS (1LBTOS) for inheritance (\figref{cntxt-impl}). 
If the access is not from the owner then  any 1LBTOS match is ignored, allowing inheritance  from 1LBTOS only if the access is the owner, 
as stated in~\secref{context}.~\figref{inheritance} shows $U3$ inheriting from $V1$, but the hardware implements only one level of inheritance for simplicity (inheritance,
a performance optimization, can be limited to any stack depth). 

\figput{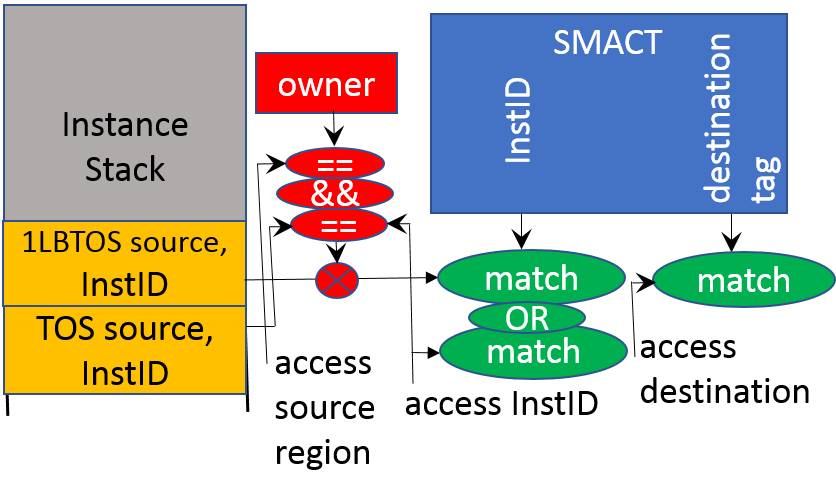}{}{Implementing instances with inheritance}

\putsubsec{free}{Revoking permissions}

{\em Whenever the sandbox boundary changes dynamically} (e.g., explicitly through memory de-allocation or recompilation, or implicitly through software access-control modification, as discussed later) -- the remaining complex behavior --
the stale permissions  must be revoked to prevent incorrectly allowing illegal, speculative accesses. Because the sandbox boundary changes too infrequently to justify cache coherence-like complexity (e.g., once per 22K-70K instructions in some SPEC benchmarks), we employ software to invalidate a multicore's SMACTs (similar to TLB shootdowns by the OS).
However, the de-allocation rate is too high for individual revocations  which incur high  overhead (e.g., 10K cycles~\cite{TLBshootdown})  mainly (1) to invoke a handler and (2) to trigger inter-processor interrupts to invoke the handler on all the cores of a multicore,  reducing overall throughput. The actual invalidation of a SMACT entry  occurs at L1 cache hit speeds (SMACT accesses) and does not impose much overhead over the usual work  in {\tt free()}. 
Consequently, we propose to amortize the handler invocation cost by {\em lazily}  batching several memory frees (our third contribution).

\begin{algorithm}
\caption{\label{alg:lazyfree}Lazy-Free}
\begin{algorithmic}[1]
\Function{LazyFree}{$ptr$}
    \State freedSize $\gets$ freedSize + findAllocSize($ptr$)
    \State count $\gets$ count + 1 \Comment Track free count
    \State pendingFree[count] $\gets$ $ptr$\Comment Append to list 
    \If {(count $>$25,000) OR (freedSize $>$2M)}  
        \State handler(pendingFree, count) \Comment Invoke handler 
    \EndIf
\EndFunction
\end{algorithmic}
\end{algorithm}

Unlike batching TLB shootdowns~\cite{LATR, TLBshootdown} which raises TLB consistency and OS semantics issues (e.g., POSIX  violation~\cite{TLBshootdown}), our
batching simply delays freeing of memory without any correctness issues.
A performance issue, however,  is that delaying the frees  increases the memory footprint and decreases locality by forcing  distant reuse of the freed memory. Less batching decreases these batching overheads but increases the above handler invocation overhead. Fortunately, frequent freeing in applications, requiring more batching,  is typically  of small chunks of memory (e.g., 64-128 B). We balance the two overheads via two thresholds based on empirically-observed rate and memory size of frees. Assuming  10K cycles per handler invocation~\cite{TLBshootdown} on each core of a 32-core multicore, we empirically find that an instruction throughput overhead of under 1\%  for our benchmarks (i.e., one invocation per 10K*32*100 = 32M cycles) requires batching up to 25K frees or 2 MB of to-be-freed memory (two thresholds). While 2 MB adds negligible memory footprint, we evaluate the batching overhead  in~\secref{main}.

We  modify {\tt free()} so that if neither threshold is exceeded {\tt free()} simply adds the to-be-freed memory to a pending set (see Alg.~\ref{alg:lazyfree}). Otherwise, {\tt free()} invokes the handler to revoke the permissions to the set, and {\em then} reclaims the corresponding  memory (i.e., adds to the free list). The handler invalidates each relevant SMACT entry using its destination. The strict ordering of invalidations followed by reclamation (and possibly re-allocation) is routine practice to avoid the well-known invalidation-reallocation race condition (e.g., in page-fault handling). 
Another option is to invalidate the entire SMACT. A corner case that may come to mind is that after a permission is revoked, a core may re-acquire the permission by accessing the freed memory. However, the permission is granted only if the access reaches commit, a  non-speculative use-after-free bug which cannot occur in correct sandboxing.  In JavaScript, for example, a dynamic down-sizing of an array would not lead to such a bug because the bounds checking would disallow access to the freed memory beyond the new array boundary. Any such access due to misspeculation of the bounds check  would not reach commit.

Analogous to data memory, the code memory can be modified or freed (e.g., dynamic linking or JIT compiling). Any new code installed in  a new region freshly acquires permissions upon commit, as usual (e.g., a new browser tab). 
However, for new code reusing an existing region (e.g., reJITing a function),
the JIT compiler must trigger a new instance in software.

To summarize,  memory allocators (64B allocation and batched lazy free with SMACT revocation), loaders (to load regions at GB alignments),  and JIT
compilers (for SMACT revocation) require changes. 

\putsubsec{ad-hoc}{Ad hoc sandboxing}

We  caution that claims about ad hoc, ``implicit'' protection schemes, where \name seemingly fails,  must {\em fully} state (a) the protection mechanism's validity for {\em non-speculative} instructions and (b) how speculation circumvents the protection. Full specification  reveals that either the schemes fail non-speculatively (i.e., without Spectre) or \name does not fail. 
Consider a {\em hypothetical}, ad hoc scheme of a simple flag protecting a secret. {\em The sandboxing system (language, compiler, runtime, or OS) -- not the application  --  must automatically check the flag before  every (pointer) access to the secret (a form of sandboxing). Without  automatic checking,  an attacker can non-speculatively  access the secret by simply omitting the  checking, if the checking is left to be voluntary by the application.} Then, the system knows (1) that the flag protects the secret and (2) about the flag's `allowed' and `disallowed' status, and can automatically revoke the permission to the secret whenever the flag changes  to the `disallowed' status (i.e., the sandbox boundary changes), {\em without} any application involvement. The flag cannot change without 
the system's knowledge, else  a malicious application can set
the flag to be `allowed'  unbeknownst to the system and {\em non-speculatively} access the secret (i.e., faulty sandbox). 
This automatic revocation upon the flag reset occurs {\em without}  memory freeing of the secret. Though used previously in {\tt free()}, the SMACT revocation is independent of {\tt free()}. 

\putsubsec{secimply}{Correctness and security analysis}

{\bf Correctness:} In the simple case, 
(1) {\em code} control flow across regions (e.g., due to shared library calls or gadgets) and (2) {\em data memory} sandbox boundary changes (e.g., memory de-allocation (\secref{free}), re-compilation (\secref{free})), and software access-control modification (\secref{ad-hoc})) are absent. {\em In this case,   the static source-destination permissions confine an attacker to its sandbox.} 
Thus, an attacker cannot speculatively access data outside its sandbox in the simple case (\tabref{threat} last column). 
Then, the only two options to access such data are: (1) reuse another region's permissions  (e.g., the victim's), or (2) reuse  the attacker's own stale permissions stemming from a sandbox boundary change  (e.g., some memory locations have moved from the attacker's region to another). These two options cover (1) inter-region control flow transfer and (2) dynamic changes of data memory sandbox boundary, respectively. 
In the first option, the attacker cannot directly reuse another
region $R$'s permissions because the instID would not match. The only way is for the attacker to transfer control flow to $R$, which must be  distinct from the attacker's region. Thus, {\em the
attacker must cross  at least its own region boundary, triggering a new instance which prevents the attacker from reusing  $R$'s existing permissions} (\tabref{threat} last column). The only exceptions are (a) a trusted owner inheriting from the caller, and (b) a caller retaining its instance upon return from a trusted owner (\tabref{call-return}).
{\em In the second option of the attacker's sandbox change, \name prevents the attacker from reusing its own stale permissions by revoking the permissions upon such change} (\tabref{threat} last column). The sandbox implementation, {\em not} the attacker, must make such change and also perform the revocation without involving the application.

\noindent
{\bf Security:} Finally, only commits insert or evict SMACT entries, so a Spectre attack on the SMACT itself is ruled out.
Further, saturating the SMACT or  the instance
counter may result in denial of service (DOS), which can be resolved by closing (killing) the offending tab (process). 
However, caches and TLBs, like the SMACT, can leak {\em non-speculative} addresses (e.g., as timing channels)~\cite{flush+reload,prime-probe} or be subjected to DOS attacks, which NDA, STT, or SDO  cannot prevent. As such, the SMACT  cannot leak any information different from what caches may.
Existing mitigations for non-speculation-based attacks~\cite{flush+reload,prime-probe} apply to \name.
Thus, \name's security is similar to that of NDA, STT, or SDO, albeit \name's DOS attack surface is larger.

\putsubsec{smact-org}{SMACT organization}

The SMACT  can be made larger or more associative as long as the SMACT latency fits within that of the D-cache.
Revoking the   permissions requires look up using the destination address (e.g., {\tt free()}'s instID is not meaningful here).
Consequently. the SMACT uses the destination to index and the InstID to match.

\figput{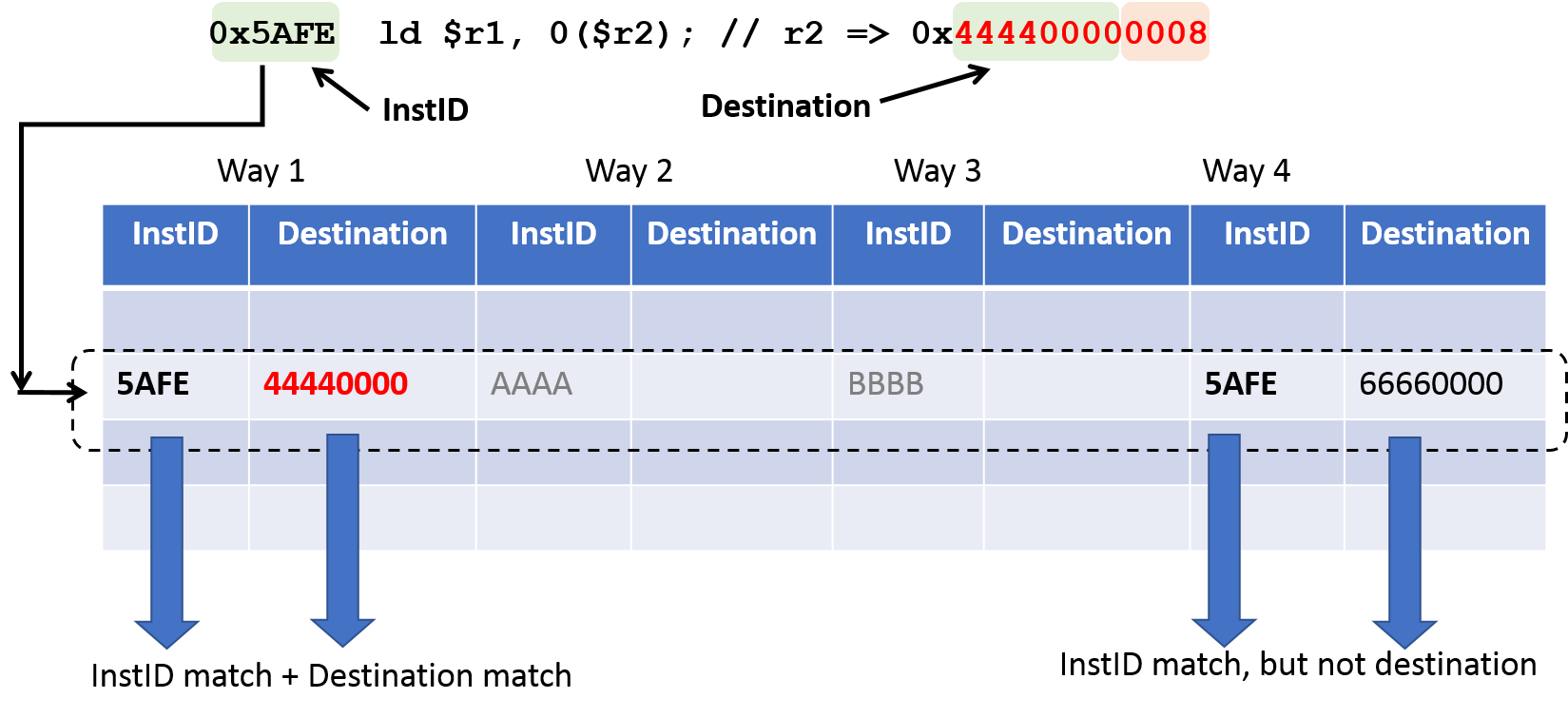}{}{SMACT with destination bit mask}

To shrink the SMACT, we observe that many upper-order destination bits of multiple entries would be the same due to locality. Accordingly, we  split the destination field into  two levels  using  bit masks (our fourth contribution). In the two-level organization,  the table  entries, which are logically the first level, correspond to {\em slabs} (e.g., 256 B). Each slab
employs a bit mask for the second level  {\em chunks} within the slab  (e.g.,  16-byte chunks within a 256-B slab result in 16 bits, as shown in \figref{bitmask}). Alternately, merging the inverted SMACT into the D-cache to  save the tag bits may be hard because the SMACT's reach is larger than the D-cache  (e.g., 256 entries with 4-KB slabs means 1 MB reach).

\putsubsec{coarsen}{SMACT destination coarsening}

Though space-efficient,  the two-level organization still adds bits whereas  coarsening entirely eliminates bits (our fifth contribution). However, naive coarsening induces  aliases which may unsafely use each others' permissions.
Consequently, coarsening the destination, unlike the source (\secref{spec-load}), needs to consider dynamic memory allocation which may arbitrarily interleave allocations for different intra-process trust domains in a fine-grained manner. Increasing the minimum allocation size for more coarsening, and hence smaller SMACT, may lead to internal fragmentation of memory and poorer cache  performance.
Per-domain private heaps would allow sufficient coarsening without fragmentation. However, multiple  domains may call the same utility functions (e.g., {\tt malloc}), so that enforcing private heaps may require changing the utility interfaces and authenticating the calling domain's identity. 
Hence, it may be hard to coarsen the destination as much as the source. 
As such, we coarsen the destination to the  minimum dynamic memory allocation granularity (e.g., 16 B in \figref{bitmask}). We increase the
 32-B minimum used in some {\tt malloc()} versions to 64 B. We account for the accompanying  performance loss (around 1\% on average).

\putsubsec{other}{Other issues}
To handle simultaneous multithreading (SMT), the SMACT may include  process identifiers (PIDs) or be privatized for each SMT context to ensure isolation. Because BTB sharing across SMT contexts can allow an attacker to induce a speculative access followed by a leak even without region crossing from the attacker to the victim, \name assumes that the BTB and branch predictor are privatized per SMT context.
To handle context switches, the SMACT  may include PIDs or be flushed at context switches (a cold SMACT has little performance impact in our context-switch-granularity simulations).
Further, we (a) clear the instance stack and (b)  either save the  instance counter leaving intact the SMACT with PIDs,  or flush the SMACT without PIDs and reset the instance counter. Similar to NDA~\cite{NDA}, we treat special registers (e.g., x86's {\tt RDMSR}) as special destinations. 
To prevent the {\em Lazy-FP-restore} attack~\cite{LazyFP}, which cannot be prevented by \name  because the secret is in the floating-point registers and not memory, we require eager FP restore.
Upon instance-counter overflows, we flush the SMACT. Finally, upon an instance stack overflow on a push for a call, the stack bottom entry is pushed out of the stack which would cause an underflow upon the return corresponding to the entry. The underflow would prevent the retention of the entry's instID, requiring a new instance upon that return.  

\putsubsec{homonym}{Homonym-based attacks}

Recall from~\secref{spectre} that the  Intel-specific homonym-based attacks exploit lazy TLB checking  or speculating past TLB misses. The unsafe transient accesses
can be prevented if the SMACT uses physical addresses instead of virtual addresses for the destination. The SMACT would be virtually-indexed and  physically-tagged using the destination address,  and accessed in parallel to the TLB and cache (i.e., the SMACT lookup critical path remains unchanged). 
The destination physical tag  from the SMACT is checked eagerly against the D-TLB output or its prediction. While D-TLB hits and predictions that are full addresses are straightforward,
partial-address predictions choose the best-matching
SMACT tag within the accessed SMACT set, effectively providing the predicted full address for the access. If the partial address yields a correct prediction in the original scheme, then the matching tag is also likely to do so. 
Any incorrect choice still results in an SMACT-permitted access and not a disallowed access, though the predicted address may be incorrect. 
In the Foreshadow-NG attack, a malicious virtual machine (VM) can update `not present' guest page table entries with  forbidden physical addresses, belonging to a victim VM, to which Intel hardware allows transient accesses under the page fault exception's shadow~\cite{foreshadow-ng}. 
The forbidden destinations may be present in the SMACT 
with the victim VM's instID but not with the malicious
VM's instID  (the malicious VM's accesses to those destinations would not ever have committed in the past). Thus, the instID mismatch prevents the attack. 
Physically-tagged structures need not be flushed upon
context switches, though TLB shootdowns have to
be applied to the physically-tagged SMACT as well.

Though our results are based on the virtual address-based design, the results hold for the  physical address-based option which does not change the SMACT's sizes, critical paths, miss rates, or miss penalties in our simulations.

\putsubsec{support}{Software support for \name}
NDA and STT hold no state, making safe and unsafe accesses indistinguishable and all accesses slow. Instead, \name uses permissions state so that most safe accesses are fast.  \name needs software support to handle the state upon inter-region code or data interaction (region alignment and revocations). However, such a need is not unusual or unreasonable. In fact,  the expectation of no software involvement set up by NDA and STT may be unrealistic. Current sandboxing requires significant software support (e.g., language and compiler guarantees). Today, trust domain-crossing is often automated (e.g., RPC stubs) where software routinely prevents non-speculative attacks (e.g., buffer overflow) to which \name's support adds prevention of speculative attacks.  \name's revocations are similar to TLB shootdowns and cache flushes upon reJITing. 
As such, with reasonable software support, \name retains simple and fast hardware.

\putsec{related}{Related Work and Key Contrast}

We discuss previous mitigations for speculation-based attacks. We have discussed speculation-based attacks in~\secref{background}.
Instead of preventing forbidden accesses, many  early hardware schemes prevent only cache-based side channels from transmitting  secrets~\cite{invisispec,delayonmiss,cond-spec,index-encrypt,DAWG}. However, non-cache side channels (e.g., AVX~\cite{netspectre},  memory~\cite{dram-channel,membus} and others~\cite{sidechannels}), some of which were discovered after the schemes, remain.

Other proposals allow the access  but block all side channels by preventing the  transmission of the secret. The schemes delay the wake-up of either (1) all (e.g., NDA~\cite{NDA} and SpecShield~\cite{spec-shield}) or (2) a subset (e.g., STT~\cite{spectaint} and SDO~\cite{data-oblivious}) of load-dependent instructions  until   the load is  either (a) ready to commit (e.g., NDA-restrictive~\cite{NDA}, STT/SDO-Futuristic~\cite{spectaint,data-oblivious}) or (b) no longer speculative (e.g., NDA-permissive~\cite{NDA}, STT-Spectre~\cite{spectaint}, SDO-Spectre~\cite{data-oblivious}). These two axes of choice lead to four bins.

In bin (1a), NDA-restrictive is somewhat comparable to \name in hardware complexity and feasibility in that both wait until commit. However, NDA-restrictive delays {\em all} load-dependent wake-ups  incurring  significant performance loss whereas \name delays only the SMACT  misses. A subtle difference is that NDA-restrictive delays only the wake-up but not the access whereas \name  delays the L1 access itself for replay later. Because \name fills
any L1 miss, irrespective of  SMACT hit or miss (\secref{spec-load}), without any delay like NDA-restrictive, \name may expose (only) the L1  latency more than NDA-restrictive. While replay is a routine feature, externally-delayed wake-up (upon reaching commit) requires changes to the issue queue. The externally-delayed wake-up would compete  for the broadcast CAM ports with normal wake-up by instructions issued in the previous cycle. In bin (2a), STT- and SDO-Futuristic avoid delaying the wake-up of  single-cycle and data-oblivious load-dependent instructions, respectively. However, this selective avoidance requires intrusive changes to (a)  the clock-critical issue queue to add a new wake-up condition and selective wake-up, and (b)  the rename logic for taint tracking. Further, STT and SDO defer memory-dependent load squashes (via the already-complex load-store queue) until possibly-multiple, matching stores become non-speculative.  STT and SDO do not address these significant complexity and feasibility challenges.

In bin (1b),  NDA-permissive requires program-order search through the reorder buffer (ROB) to detect which, if any, later instructions have  become non-speculative when a branch is resolved (i.e., all previous branches have been resolved). 
The ROB is a no-search, true FIFO holding around 200 instructions, whereas, for reference, the associatively-searched issue queue holds only around 50 instructions. Such large search incurs either latency 
(e.g., 4-cycle sequential, partitioned search) or power (1-cycle, brute-force, parallel search), and  energy either way.
STT- and  SDO-Spectre in bin (2b) remain more complex and less feasible than NDA-permissive in bin (1b) because of the above changes to the issue queue and rename logic {\em in addition to} searching the ROB.
In contrast, \name employs simple table look-ups and replay upon reaching commit.

Compiler-based mitigations limit speculation~\cite{intel-spectre-whitepaper}, or include branches~\cite{v8-mips-revert} or conditional instructions~\cite{v8-masking-blog,ldh-googledoc,old-csdb-spec,new-csdb-spec} which themselves may be predicted~\cite{sigarch-blog}. Moreover, some of the mitigations have uncertain semantics due to undocumented features~\cite{intel-spectre-whitepaper} or changed specifications~\cite{old-csdb-spec,new-csdb-spec}. Chrome's \textit{site isolation} places browser tabs in different processes so virtual memory disallows forbidden accesses~\cite{google-site-isolation}, but incurs performance loss~\cite{sabc} and is not adopted by other browsers.
VirtualGhost~\cite{vghost} blocks certain shared page-based Spectre-v1 attacks but does not address bounds checking.

\putsubsec{privacy}{Privacy}
Illegally accessing secrets and leaking ({\em security}) is { \em fundamentally } different from legally holding secrets and leaking ({\em privacy})  (Sec.IV second-last paragraph). SafeBet addresses the former. Building on STT, Speculative Privacy Tracking (SPT)~\cite{spt} addresses both.
SPT aims to prevent speculatively transmitting non-speculatively accessed data, a privacy problem, similar  to {\em NDA-strict} where (committed) registers can be leaked (\secref{exclusions}). For the security problem, SPT retains
STT's functionality of preventing the transmission of  speculatively-accessed data.  

In SPT,  a transmitter becoming non-speculative allows other, {\em statically-different}, speculative transmitters of the {\em same value} to proceed - a novel application of the privacy notion of declassification. However, the first transmitter in every dynamic instance of the producer-transmitter code stalls until becoming non-speculative. 
In \name, by contrast, a committed source-destination pair enables other speculative {\em dynamic instances} of the {\em statically-same pair} to proceed, so that only SMACT misses cause stalls. 
As such, SPT (1) {\em does not exploit the difference between the  two problems} and (2) inherits STT’s significant hardware complexity and feasibility challenges, as discussed above.
For reasonable hardware, SPT (and STT)  must wait until commit and incur significant slowdowns (worse than {\em NDA-permissive-4} in~\figref{main}).

To ensure privacy, secret-handling functions (e.g., constant-time cryptography and other password-handling routines) carefully prevent non-speculative leaks (NSLs) by avoiding secret-dependent branches, memory accesses and variable-latency operations. 
Instead of SPT, the functions can avoid speculative leaks {\em in software} by {\em disabling}  (a) speculation and (b) SMACT {\em insertion} of new permissions, from the function start (set a  status bit)  until end (clear the bit) including {\em all} in-between {\em dynamic} instructions, By squashing fetch-decode upon decoding a branch and stalling fetch until the branch resolves, part (a) prevents speculative {\em leak} of registers {\em within} the function (SPT’s improbable, {\em NDA-strict}-like threat model without a  proof-of-concept implementation). Part (b) prevents sticky speculative-access permissions {\em after} (and redundantly within) the function. Even with disabled insertion, the SMACT {\em always} checks accesses, so {\em any untrusted or trusted} code’s  adversarial or accidental disabling can affect only performance, {\em not} security. Part (a) is unnecessary if leaks must speculatively load the secret instead of directly leaking registers, a more reasonable  model.

The disabling poses far lower programmer-burden than the algorithm/semantics-dependent, unautomated NSL-avoidance, and  is done {\em only} when secrets are accessed which is known (else NSLs cannot be avoided).  Because NSL-avoidance already incurs slowdowns, such functions may naturally be run infrequently. The claim that all data is secret requiring the disabling always, not only in secret-handling  functions, is invalid because then {\em NSLs would have to be avoided always} incurring infeasible programmer burden and slowdown. In {\em fundamental} contrast, security must be ensured {\em always} (i.e., {\em any} access may steal secrets)  but requires only easily-predictable,  low-overhead, compiler-inserted bounds-check branches.

In summary, we advocate (i) using SafeBet for always ensuring  security and (ii) non-leaky code with the disabling  for privacy (wait until  commit {\em only} when secrets are accessed, {\em not unnecessarily always} like SPT). SafeBet makes the {\em predominantly–common, no-attack} case fast, whereas SPT makes {\em most} cases slow. 

\tabput{param}{
\small
\begin{tabular}{|r|p{2.6in}|}
\multicolumn{2}{c}{\bf Baseline CPU}\\
\hline
Core  & 2 GHz, 8-wide, 64-entry issueQ, 192-entry ROB \\  
L1 D/I & 32 KB, 8-way associative, 64B block\\
L2 & 256 KB 16-way associative, 64B block\\
L3 slice & 2 MB, 16-way associative, 64B block\\
\hline
\multicolumn{2}{c}{\bf Safebet (table size include tag and data)}\\
\hline
SMACT & 512 entry, 8-way, 4 KB slab, 64B chunk, 8.3 KB\\
\hline
\end{tabular}

\vspace{-0.2in}
}
{Baseline CPU and Safebet parameters}

\putsec{method}{Evaluation Methodology}

We implement \name  in {\em gem5}, a software simulator for performance, under syscall emulation~\cite{GEM5}.  We use McPAT~\cite{mcpat} for evaluating energy using 22 nm technology.

\tabputW{benchmarks}{
\footnotesize
\begin{stripetabular}{|r|p{0.25in}|c|c|c|c|c|c|c|c|c|c|c|c|c|c|c|c|c|c|}
\hline
\multicolumn{2}{|r|}{\bf Benchmark} & 
\rot{exchange2\_17}& 
\rot{astar\_06}& 
\rot{leela\_17}& 
\rot{namd\_06}& 
\rot{fotonik3d\_17}& 
\rot{sphinx3\_06}& 
\rot{nab\_17}& 
\rot{deepsjeng\_17}& 
\rot{xz\_17}& 
\rot{imagick\_17}& 
\rot{perlbench\_17}& 
\rot{cactuBSSN\_17}& 
\rot{milc\_06}& 
\rot{gcc\_06}& 
\rot{bwaves\_17}& 
\rot{mcf\_17}& 
\rot{leslie3d\_06}& 
\rot{lbm\_17}
\\
\hline 
\multicolumn{2}{|l|}{Base IPC} & 1.8 & 0.8 & 1.0 & 2.6 & 1.9 & 1.8 & 1.9 & 1.4 & 1.2 & 1.5 & 1.1 & 1.9 & 1.4 & 1.0 & 0.7 & 0.6 & 0.5 & 0.3\\
\multicolumn{2}{|l|}{L3 MPKI}& * & * & * & * & * & * & * & * & * & 0.1 & 0.2 & 0.2 & 0.3 & 0.7 & 1.5 & 3.3 & 4.2 & 6.5\\ 
\hline
\hline

& Dst.Sl & * & * & 0.1 & * & * & * & * & 0.6 & 0.2 & 16.8 & 0.5 & * & 0.8 & 0.3 & * & 12.4 & 1.5 & 4.2\\
& Dst.Ch & 2.9 & 0.2 & 0.5 & 0.1 & 0.1 & 2.5 & 0.1 & 0.2 & 0.6 & 4.0 & 2.2 & 2.4 & 0.8 & 5.8 & 20.0 & 74.6 & 67.1 & 133\\

& Instance & 4.8 & * & 0.5 & * & 2.3 & 5.4 & 3.5 & * & * & 8.2 & 0.3 & 6.8 & 21.2 & * & 17.8 & * & * & *\\
\multirowcell{-4}{ 
\rot{smact mpki}}& Total & 7.6 & 0.2 & 1.0 & 0.1 & 2.3 & 7.9 & 3.6 & 0.8 & 0.9 & 29.0 & 3.1 & 9.2 & 22.8 & 6.1 & 37.9 & 87.0 & 68.6 & 137 \\

\hline
\end{stripetabular}
\vspace{-0.2in}
}
{Benchmark (* means $<$ 0.1 and Dst.Sl and Dst.Ch stand for destination slab and chunk)}

\noindent
{\bf Systems: } 
We compare NDA (permissive and restrictive) and \name, 
where the latter two are comparable in security and  hardware complexity.
As discussed in~\secref{related}, the other previous work  
block transmission through only some but not all side channels (i.e., less secure than \name), or incur significantly higher hardware complexity than \name to plug all side channels. We configure {\em gem5} with typical CPU parameters, as listed in~\tabref{param}. Because we run sequential benchmarks we consider one core slice of a typical multicore. Based on the SMACT parameters, 
the  64-bit destination address results in 12 bits of slab offset, 6 bits of index, and 46  bits of tag.
Each SMACT entry holds a 64-bit  destination bit mask and 22-bit instance (\figref{bitmask}). SMACT's  size is 2,944 B for tag + 5504 B  for data = 8.3 KB. 

\noindent
{\bf Benchmarks:}  We use a mix of  SPEC'17 (speed) and SPEC'06 CPU benchmarks, as listed in~\tabref{benchmarks}. The rest of the benchmarks either do not run on unmodified {\em gem5} under {\em Syscall Emulation} or are present in both suites in which case we use the SPEC'17 version. Each benchmark  has two regions: the application and libraries, which trigger \name's instance annotation (\secref{context}). We skip 1 billion instructions and run 500 million with warmed-up caches. 

\noindent
{\bf Malloc/free:}  We modify {\tt glibc 2.17} with default {\em malloc-32()} to incorporate our {\em lazy-free()} (\secref{free}) and {\em malloc-64()} (\secref{coarsen}).
Because the 500M-instruction {\em gem5} samples may not capture the relatively infrequent {\em free()}, we measure the performance impact of these changes on real hardware using many full-benchmark runs (2.8 GHz, 4-core Opteron 6320 with 64-KB L1, 512-KB L2 and 8-MB shared L3). We use Pin~\cite{pin} to count the number of frees per million instructions and the amount of memory freed.

\putsec{results}{Results}

We (1) compare the performance of \name and NDA, (2)  show  \name's energy impact, (3) isolate the impact of \name's two-level bit mask representation, source coarsening, and permission inheritance, and finally, (4) evaluate \name's sensitivity to the SMACT sizes. 

\figputW{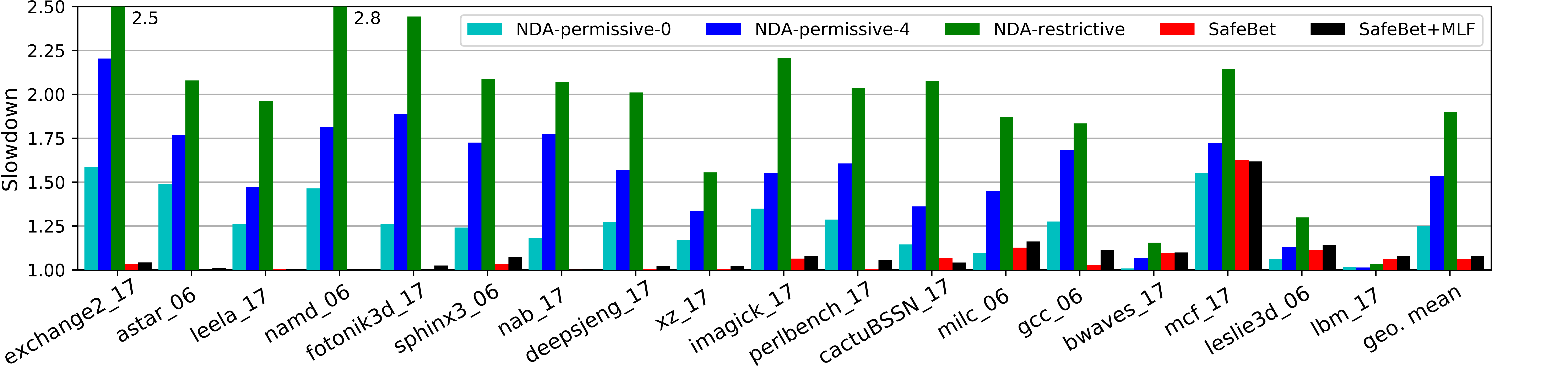}{}{Performance Comparison}

\putsubsec{main}{Performance}
\figref{main} shows the execution times,  normalized to those of the insecure baseline  (Y-axis),  for NDA-permissive with 
zero-cycle delay for dependent wake-up ({\em NDA-permissive-0}), NDA-permissive with four-cycle delay for dependent wake-up  ({\em NDA-permissive-4}), NDA-restrictive, \name, and \name including the 
overheads of {\em malloc-64} and {\em lazy-free} ({\em \name+MLF}).
On the X-axis, the benchmarks are arranged by increasing  L3 miss rate (\tabref{benchmarks})  for showing performance trends clearly. 

\figput{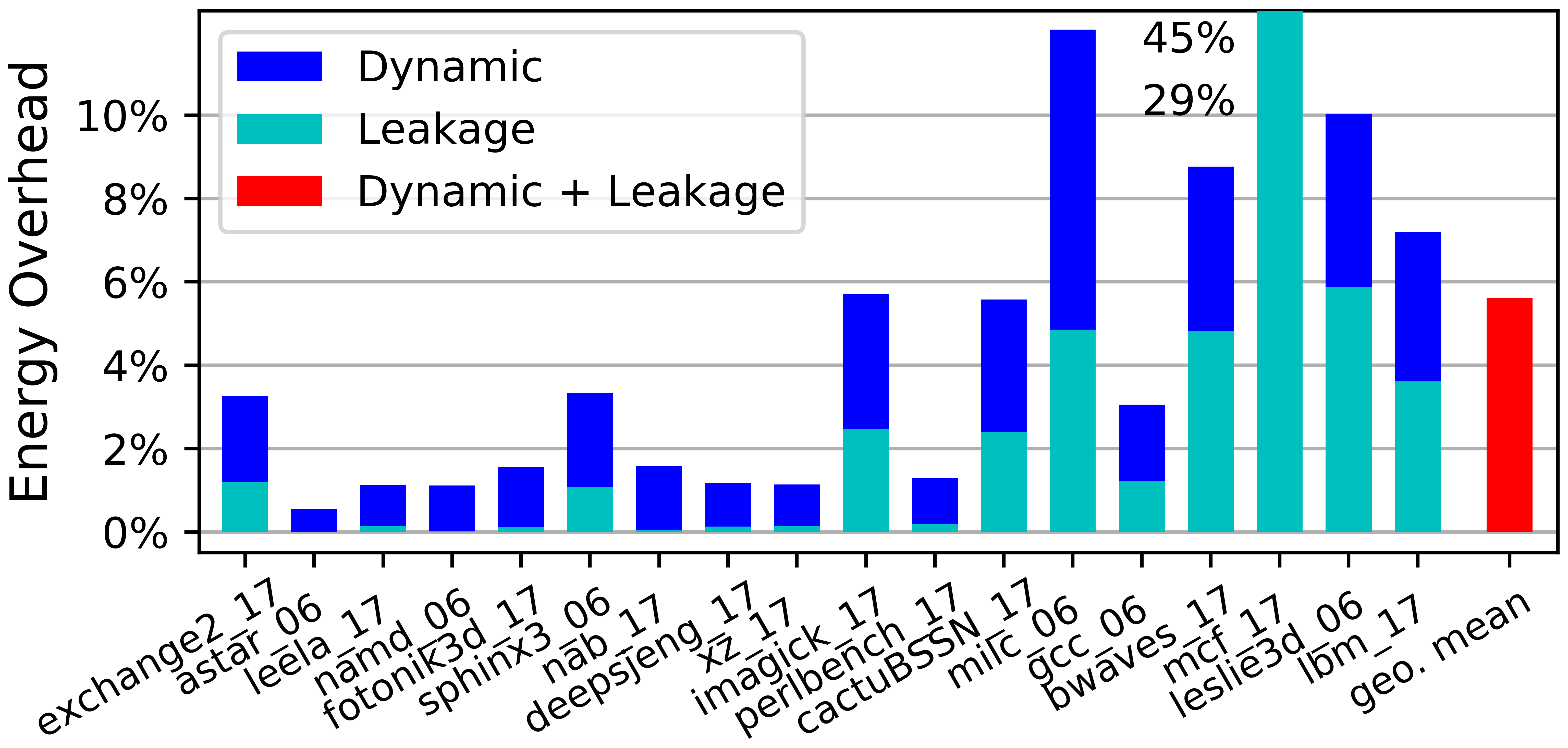}{}{Energy Overhead}

By delaying the dependent wake-up until {\em each} load becomes non-speculative,  NDA-permissive-0, despite assuming zero-cycle ROB search (\secref{related}), incurs significant slowdowns across the board (25\% on average). As discussed in~\secref{spec-load}, such delay degrades instruction- and memory-level parallelism (ILP and MLP). As expected, NDA-permissive-4, which assumes slower but still aggressive four-cycle ROB search,  incurs even more slowdowns. NDA-restrictive, which is somewhat comparable to \name in hardware complexity (\secref{related}), delays the wake-up until each load reaches commit, incurring severe degradation. In contrast, by replaying upon reaching commit only the SMACT misses, \name  incurs negligible to modest slowdowns (6\% on average),  except for {\em mcf\_17}.  It is worth noting that \name achieves such good performance with a reasonably-sized SMACT (8.3 KB for tag and data).

Finally, adding {\em malloc-64} and {\em lazy-free} overheads only slightly worsens \name's slowdowns.  {\em perlbench\_17} and {\em gcc\_06} have
free rates of  335 and 18 per million instructions and average sizes of 67 B and 1.87 KB  per free, respectively, while the other benchmarks' rates are at least 10x, and often 100-1000x, lower than that of {\em gcc\_06}. Consequently, only these two benchmarks have noticeable impacts of 5\% and 8\% slowdowns, respectively.

\figputW{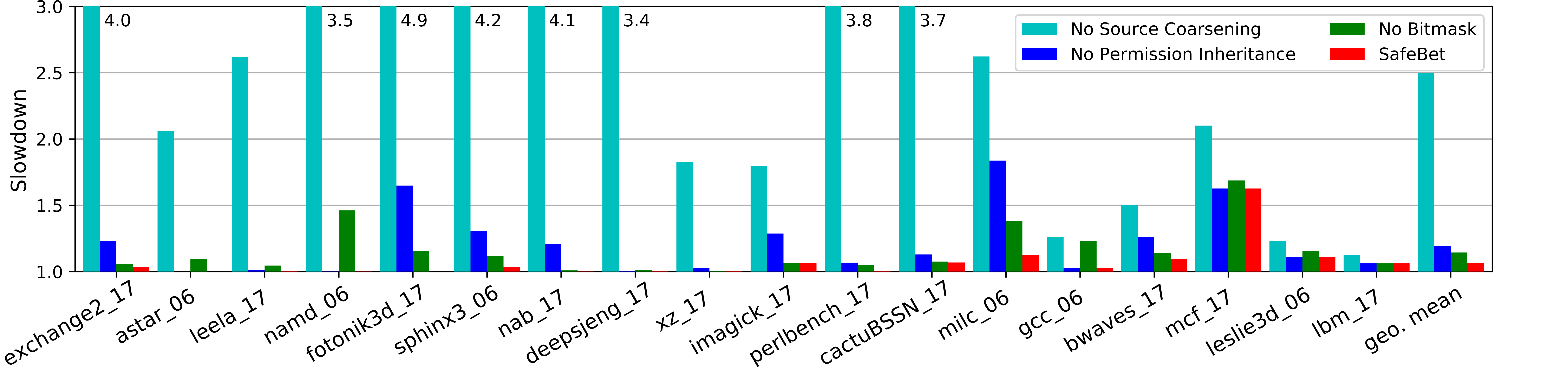}{}{Isolating \name's techniques}

To understand \name's slowdowns across the benchmarks, we identify three factors: the SMACT miss rate, and the baseline's L3 miss rate and instructions per cycle (IPC), shown in~\tabref{benchmarks}. In general, we expect the SMACT miss rate to track closely the L3 miss rate. Clearly, a low SMACT miss rate implies little performance loss. The benchmarks to the left of {\em xz\_17} in~\figref{main} fall in this group. 
A modest SMACT miss rate combined with low L3 miss rate and high IPC implies that the SMACT misses incur only a short wait until commit resulting in only a modest performance loss. The benchmarks
between {\em imagick\_17} and {\em milc\_06} in~\figref{main}  fall in this class. Specifically,  {\em imagick\_17} has high SMACT miss rate and  {\em cactuBSSN\_17} has little slowdown. {\em cactuBSSN\_17}'s high instruction-level parallelism (ILP), attested by its high baseline IPC despite a modest L3 miss rate (\tabref{benchmarks}), overlaps \name's delay until commit. Similarly, {\em imagick\_17}'s high SMACT miss rate is offset by its high ILP.

However, a moderate to high SMACT miss rate combined with moderate to high L3 misses and moderate to high IPC implies  high ILP  and memory-level parallelism (MLP) and that many L1 miss prefetches (\secref{spec-load}) would complete  {\em before} the load reaches commit. The benchmarks between {\em gcc\_06} and 
{\em leslie3d\_06} fall in this class where any SMACT miss latency for waiting until commit is exposed incurring performance loss. 
{\em mcf\_17}, the most extreme case in this class,  has reasonable MLP, achieving a reasonable baseline IPC despite high L3 miss rates. Consequently, {\em mcf\_17}'s high SMACT miss rate 
implies long exposed wait until commit  and high slowdown.
\name is slower than NDA-permissive-0 because the former's replay further delays the accesses whereas the latter delays only the dependent wake-up but not the access itself (\secref{related}). With a full 16-MB L3 instead of a 2-MB L3 multicore slice,  \name's slowdown for {\em mcf\_17} drops to 18\% (i.e., fewer L3 misses means higher IPC and shorter wait until commit).

Finally, a high SMACT miss rate, high L3 miss rate and low IPC imply low instruction- and memory-level parallelism so that most L1 miss prefetches complete {\em after} the load reaches commit. {\em lbm\_17} falls in this group where any waiting until commit due to SMACT misses is hidden under the L1 miss resulting in little performance loss.

\tabref{benchmarks} breaks down the SMACT  misses into destination slab misses ({\em Dst.Sl}, destination chunk misses when the slab is present ({\em Dst.Ch}), 
and finally instance misses when there is an instance mismatch (despite permission inheritance). In general, the SMACT miss rates track the L3 miss rates in intensity, with three exceptions. 
{\em imagick\_17} has weak SMACT slab-level spatial locality though only modest L3 miss rate.  {\em bwaves\_17} has the opposite behavior wherein its slab misses are fewer than chunk misses due to spatial locality, but a slab eviction removes all its chunks magnifying the chunk miss rate. 
Finally, {\em milc\_06} and {\em bwaves\_17} see significant instance-misses despite inheritance.

\putsubsec{energy}{Energy}
\figref{energy} shows \name's energy overhead, broken up into leakage and dynamic components, over the insecure baseline considering the core, L1, L2, and L3 slice. While \name's tables are too small to impact the total energy, \name's overhead comes from increasing dynamic energy due to replays and leakage due to the slowdowns. The overhead ranges from negligible for low SMACT miss-rate benchmarks (under 2\%) to modest for high SMACT miss-rate benchmarks (5\%-12\%) except  {\em mcf\_17}, where the baseline leakage energy is high (45\%) due to its low IPC. This high leakage combined with the 
large slowdown leads to high leakage overhead (31\% of baseline total energy). 
Moreover, {\em mcf\_17}'s high SMACT miss rate causes many replays, increasing dynamic energy. {\em milc\_06} and {\em leslie3d\_06} are similar but to a much lesser extent.
Additionally, \name's area overhead is under 0.5\%.

\putsubsec{isolation}{Isolating \name's techniques}
\figref{isolations} isolates the impact of the SMACT's bit mask representations, source coarsening, and permission inheritance. Because the SMACT simply uses the minimum {\tt malloc} allocation, we do not isolate destination coarsening. 
The Y-axis shows execution times normalized to those of 
\name without  (a) bit masks (i.e., 64-B slabs),  (b) source coarsening (i.e., instruction granularity instead of region granularity), (c) permission inheritance, and (d) full \name for reference.  

\name without SMACT bit masks spends extra SMACT entries for chunks within a slab increasing the pressure on the SMACT, whereas the bit mask captures such chunks in one entry. In general, higher ratios of destination chunk miss rate to destination slab miss rate signify slab-level locality.
In benchmarks with significant locality, the elimination of destination bitmasks places pressure on SMACT capacity, resulting in higher chunk miss rates (\tabref{benchmarks}). The additional SMACT misses cause slowdowns (see \figref{main}, e.g., {\em namd\_06}, {\em milc\_06}, {\em gcc\_06}, and {\em mcf\_17}).
The absence of source coarsening leads to severe locality splintering -- the same destination requires  a distinct permission for each instruction within the same region (\secref{spec-load}) --- which causes significant SMACT misses and slowdowns. All benchmarks, incur high SMACT miss rates (10s to 100s MPKI) causing significant slowdowns due to severe loss of ILP and MLP, and high cache bandwidth pressure due to excessive replay. These slowdowns are higher than those of NDA-permissive which does not incur this pressure (\figref{main}).

Finally, the lack of permission inheritance leads to unnecessary SMACT misses in the callee, and hence slowdowns (e.g., {\em exchange2\_17, fotonik3d\_17, imagick\_17, and milc\_06}). 

\figput{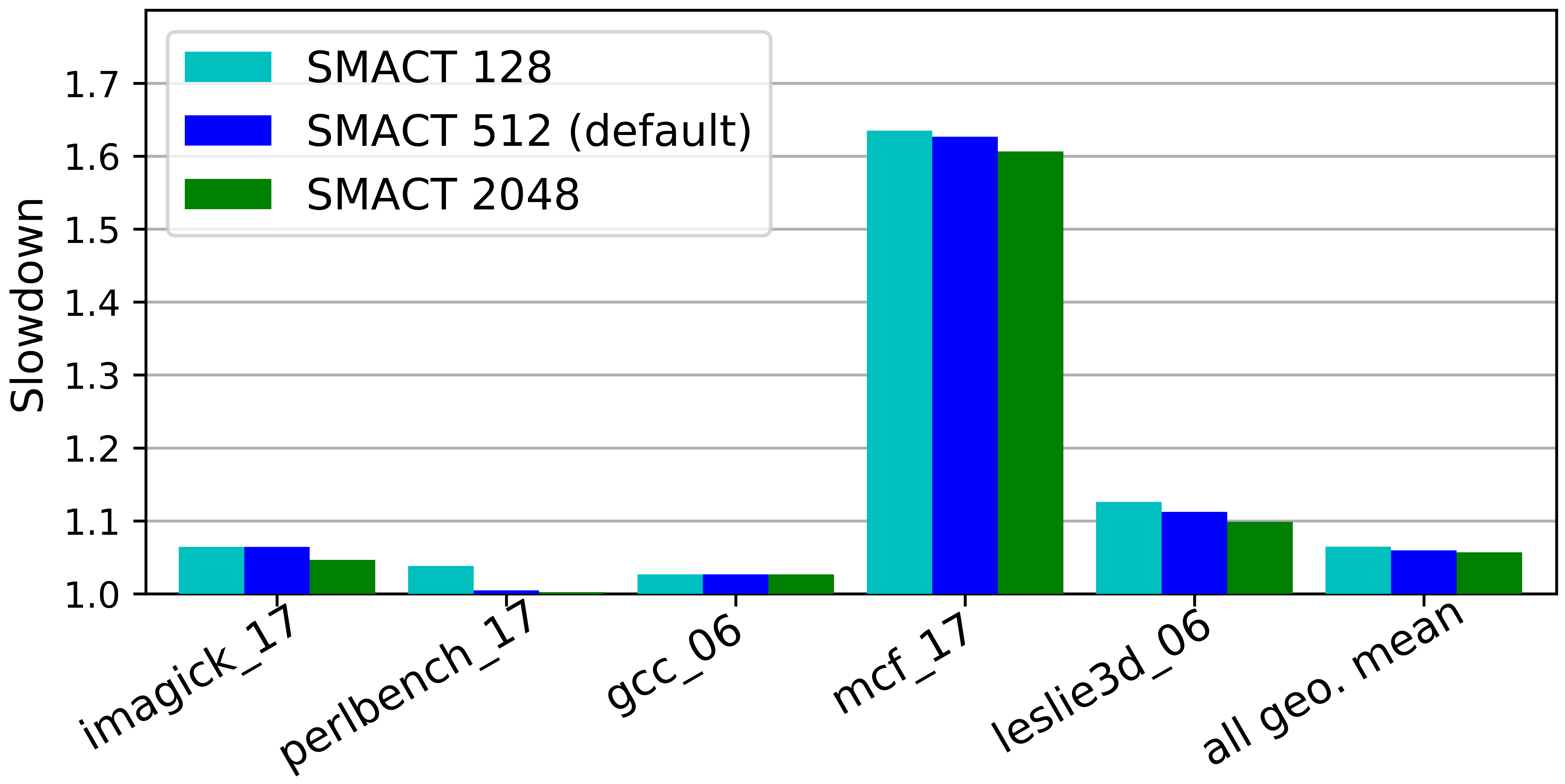}{}{Sensitivity to SMACT size}

\putsubsec{sensitivity}{Sensitivity}
\figref{size_sensitivity} varies the SMACT size as 128, 512 (default), and, 2K entries.
For each set, the  X-axis shows a subset of some interesting benchmarks  and the geometric mean of {\em all} the benchmarks. Across the SMACT sizes,  performance varies little for these benchmarks which incur some slowdown (\figref{main}) or modest to high SMACT miss rates (\tabref{benchmarks}). The other benchmarks (not shown) have virtually no change because the SMACT reach even for 128 entries (128 x 4-KB slab = 512 KB reach) is sufficient for these benchmarks. However, a 4-entry SMACT severely degrades many benchmarks (not shown). On the other side,  {\em mcf\_17} needs 8K entries to approach the baseline (not shown); the 2K-entry SMACT is not significantly better than the default SMACT size. Varying the SMACT  slab  size as 2-8 KB produces little change. These results are not shown. 

\putsec{conclu}{Conclusion}
Variants of Spectre and Meltdown pose a serious security challenge arising from unsafe speculative accesses that can penetrate trust boundaries.
Previous work allows potentially-unsafe speculative accesses and then prevents the secret's transmission by delaying all or many of the access-dependent instructions {\em even in the predominantly-common, no-attack case}, incurring performance loss or hardware complexity. 
Instead, \name  allows {\em only}, and {\em in the common case} does not stall {\em most}, safe accesses
based on the key observation that in the absence of (i) {\em code} control flow transfer across trust domains or (ii) {\em data memory} sandbox boundary changes, a speculative access is safe if the location's access by the same instruction has been committed previously; otherwise,  the speculative access is potentially unsafe and must wait until reaching commit.  
\name tracks such 'non-speculative source instruction code region-destination location' permission pairs in the {\em speculative memory access control table (SMACT)}  which guarantees no false positives (i.e., unsafe accesses are never treated as safe) and enables high performance because the common-case safe speculative accesses are not delayed. To prevent one dynamic execution context from reusing the permissions of other contexts via control flow transfer across trust domains, \name creates a new dynamic instance for each such transfer and replaces the static source in the permissions with the instance identifiers.  To prevent 
the use of permissions that are stale due to data memory sandbox boundary changes, \name revokes such stale permissions efficiently in software. With virtually no change to the processor pipeline, \name prevents all variants of Spectre and Meltdown except Lazy-FP-restore, 
using any current or future side channel.
Simulations show that \name and {\em NDA-restrictive}, which are comparable in security and hardware complexity, are slower on average than the unsafe baseline by 6\% and 83\%, respectively.

\bibliographystyle{plain}
\bibliography{local}

\end{document}